\documentclass[%
reprint,                     %onecolumn,
amsmath,amssymb,
aps,
]{revtex4-2}

\usepackage{graphicx}% Include figure files
\usepackage{dcolumn}% Align table columns on decimal point
\usepackage{bm}% bold math
\usepackage{times}
\usepackage{hyperref}

\usepackage{comment}
\begin{document}
	
	\preprint{APS/123-QED}
	
	\title{ All non-locally Realized Continuous Variable Bipartite Gaussian States are Entangled
	}
	
	\author{Souvik Agasti}
	\email{souvik.agasti@uhasselt.be}
	
	\affiliation{
		IMOMEC division, IMEC, Wetenschapspark 1, B-3590 Diepenbeek, Belgium
	}%
	\affiliation{
		Institute for Materials Research (IMO), Hasselt University,	Wetenschapspark 1, B-3590 Diepenbeek, Belgium
	}%

	
	\begin{abstract}

		We investigate the connection between entanglement and non-locality between continuous-variable bipartite Gaussian states. The investigation initiates with formulating non-locality by using the phase-space Wigner representation of Bell's function. Furthermore, our analysis shows entanglement to be necessary for nonlocality, but not sufficient for it; however, nonlocality is sufficient to ensure entanglement.

	\end{abstract}
	
	\maketitle
	
	%\tableofcontents

	\section{Introduction}\label{Introduction}

	Entanglement is a fundamental aspect of quantum mechanics, had remains to be the key feature in the development of the modern-day advanced quantum technologies, including quantum information processing \cite{entanglement_quantum_cryptography}, quantum computation \cite{quantum_computation}, and communication \cite{entanglement_swapping}.  Gaussian states have drawn attention for the generation and application of continuous variable (CV) entangled states \cite{quantum_teleportation, Entanglement_CV_Plenio}. A popular example of such Gaussian CV state is the two-mode squeezed vacuum (TMSV) state which show entanglement between two parties: signal(S) and idler(I) modes; had remained in the intensive interest of the community due to its application in sensing and manipulation of quantum states, e.g. quantum memory systems \cite{Quantum_memory}  and even in gravitational wave metrology \cite{study4roadmap, Polzik_GW_Spin_PRD, my_BAE}.
	
	%  hybrid

	Entanglement represents a correlation between parties in hybrid systems, therefore playing the key role in the theoretical establishment of quantum mechanics. 	The entanglement criteria of a bipartite system are dependent on separability limits. The separability exists if and only if the bipartite quantum state $(\rho)$ can be expressed in the form of $\rho = \sum_j p^j \rho^j_I\otimes\rho_S^j$ where $\rho^j_I,\rho_S^j$ are the normalized states of of the parties I and S, respectively, and the probability $p^j\geq 0$ satisfies $\sum_j p^j = 1$. The separability limit was first proposed by Peres \cite{Peres_entanglement} and afterwards by Horodecki for higher-dimensional cases \cite{Horodecki_entanglement}. In the Heisenberg picture, the idea has furthermore been established and formulated by Simon \cite{Simon} and Duan et. al. \cite{Duan_Inseparability_entanglement}. Many other entanglement measures have also been used in different circumstances, e.g von Neumann entropy of reduced density operator, maximally optimized hybrid squeezed quadrature, Cauchy–Schwarz inequality, Hillery–Zubairy criterion, GHZ (Greenberger–Horne–Zeilinger) equality \cite{ent_nonL_rev}.

	The impact of entanglement reflects a profound feature of quantum measurements on bipartite quantum systems, which is non-local realism, which solved the Einstein-Podolsky-Rosen (EPR) measurement paradox \cite{EPR_original, EPR_bell}. The test of a hidden variable that puts a limitation on local realism has been determined by Bell’s inequality \cite{TMSV_nonlocal_Bell_Wigner, bell_homodyne_1, bell_homodyne_2}. Quantum
	nonlocality has been tested by exploring several types of Bell’s inequalities, which are experimentally realizable through homodyne measurements \cite{bell_homodyne_1, bell_homodyne_2}.  Such a type of inequality was first formulated by Clauser, Horne, Shimony, and Holt (CHSH) \cite{bell_CHSH, bell_CH}, and further, it is expressed in phase space using Wigner quasiprobability functions \cite{Banaszek_TMSV_bell, Banaszek_TMSV_bell_PRL}.
	
	Even though many theoretical conditions have been prescribed to witness entanglement and eventually quantify it, there is no direct analytical expression available for the quantification of non-locality, especially in the Heisenberg picture, which is still a gray area to explore. The maximization of Bell's function has usually been numerically determined so far, while analyzing homodyne measurements, as the quantification of non-locality \cite{TMSV_nonlocal_Bell_Wigner, mypaper_TMSV_filter_STD}. Therefore, it becomes essential to provide an analytical expression that formulates Bell's inequality, especially for Gaussian states, which are often used in quantum optical experiments.

	Although entanglement and nonlocality are different phenomena, the latter has been vaguely considered as evidence of entanglement \cite{ent_nonL_rev}. However, it has been realized that entanglement can be witnessed even though systems are locally realizable \cite{mixed_entanglement_nolocal, mypaper_TMSV_filter, mypaper_TMSV_filter_STD}. In contrast, strong quantum nonlocality has also been realized without entanglement as well \cite{Quantum_Nonlocality_without_Entanglement}, and as a consequence, a limited class of states nonlocally realizable can witness quantum entanglement \cite{Buscemi_Nonlocality_Entanglement}. In fact, all classically correlated and non-product states admit violation of Bell's inequality \cite{Werner_classical_bell, Gisin_Bell, Doherty_bell}. However, Barrett has proven that	entangled mixed states do not always violate a Bell inequality \cite{Barrett_bell}. Therefore, it remains an open quension and becomes necessary to see how the non-local behavior of Gaussian CV bipartite states is being modified with entanglement, and are there any relation can be established between the two phenomena.
	
	In this article, from the correlation between two parties for CV Gaussian states, using Wigner phase-space formalism, we propose a formulation of maximized Bell's function as a measure of non-locality%, and the boundary condition, prescribes a non-locality criterion
	. Furthermore, we prove that entanglement could be necessary, but may not be sufficient, for non-locality; however, the nonlocality is sufficient for entanglement for bipartite Gaussian states.

	\section{Correlation Matrix in Standard Form}
	
	The correlation between two parties in Gaussian states can be completely determined by a $4\times 4$ real symmetric correlation matrix:
	
	\begin{equation}\label{corr_mat_original}
		V' = \begin{bmatrix}
			V'_I & V'_{IS} \\
			{V'_{IS}}^T & V'_S
		\end{bmatrix},
	\end{equation}
	where $V'_I(V_S)$ and $V'_{IS}$ are the $2\times 2$ real matrices, represent for the systems $I(S)$ and their cross-correlation, respectively. While studying their separability property, it remained convenient to transform the correlation matrix to some standard form using a local linear unitary Bogoliubov operator (LLUBO) $U_l = U_I \otimes U_S$ \cite{Duan_Inseparability_entanglement, Simon}, and we will use the same standard form to formulate non-locality as well. 
	%In Heisenberg picture, each matrixes of $U_l$ satisfies $U_k^T U_k^\dagger = H_k^T (k\in[I,S])$, where $H_k$ is a $2 \times 2$ real matrix which satisfies $\det (H) = 1$.
	The LLUBO operation works with a combination of squeezing transformation and rotation \cite{LLUBO_transformation}, which transforms the correlation matrix of Eq. \eqref{corr_mat_original} to a standard form as
	
	\begin{equation}\label{corr_mat_transformed}
		V = \begin{bmatrix}
			n & & c_1 & \\
			& n &  & c_2\\
			c_1 &  & m &  \\
			& c_2 & & m 
		\end{bmatrix} \, , \,  \, (n,m\geq \frac{1}{2})
	\end{equation}
	
	The operation goes with diagonalizing $V'_I$ and $V'_S$ by the first LLUBO operation using local orthogonal $U_{I,S}$, and then squeezing locally to obtain $V_I = nI_2$ and $V_S = mI_2$ where $I_2 = \text{diag} [1, 1] $. After these steps of operation, the transformed version of $V'_{IS}$ is furthermore diagonalized by another LLUBO with a suitable choice of orthogonal $U_{I,S}$, which does not influence $V_I$ and $V_S$. The newly generated parameters in Eq. \eqref{corr_mat_transformed} are related to the inverients as $\det(V'_I) = n^2, \det(V'_S) = m^2, \det(V'_{IS}) = c_1c_2$ and $\det(V') = (nm-c_1^2)(nm-c_2^2) $. For a bipartite Gaussian state, the mixedness of the state is quantified by $\text{Tr} (\rho^2) = \frac{1}{4 \sqrt{\det(V )} }$, which leads to 
	
	\begin{equation} \label{mixedness}
		4(nm-c_1^2)(nm-c_2^2) \geq \frac{1}{4} 
	\end{equation}
	
	The inequality furthermore gives: $ nm - |c_1c_2| \geq \frac{1}{4}$. 
	
	\section{Quantum non-Locality}
	
	The non-locality is justified by the violation of Bell’s inequality, which can be quantified by the maximum value of the Bell’s function $(|B_{max}|)$, expressed in terms of Wigner functions as \cite{TMSV_nonlocal_Bell_Wigner, Banaszek_TMSV_bell_PRL, Banaszek_TMSV_bell} 
	
	\begin{equation}\label{bell_func_deff}
		B = \pi^2 \left[ W(u_{00}) + W(u_{I0}) + W(u_{0S}) - W(u_{IS}) \right]
	\end{equation}
	
	where 
	
	\begin{equation}
		W( {u_{IS}}) = \frac{1}{ 4 \pi^2 \sqrt{\det(V) }} \exp \left[ -\frac{1}{2} {u_{IS}}^T {V}^{-1} {u_{IS}}\right]
	\end{equation}
	
	represents the Wigner function for the vector ${u_{IS}} = [Q_I ,P_I , Q_S, P_S]^T$ in phase space $([Q_0 ,P_0] = [0,0])$. The sufficient condition for non-local realization of a quantum mechanical bipartite state is the violation of $|B_{max}| \leq 2$, and a larger $|B_{max}|$ indicates stronger non-locality. The Bell function defined in Eq. \eqref{bell_func_deff} will be maximized when the following function at the exponent will be maximized

	\begin{align}
		& -\frac{1}{2} {u_{IS}}^T {V }^{-1} {u_{IS}} = -\frac{1}{2} \big[ m \alpha_I^2  +n \alpha_S^2  \\
		& -2 c \alpha_I \alpha_S (\cos\phi \cos\theta_I \cos\theta_S + \sin\phi \sin\theta_I \sin\theta_S) \big] \nonumber
	\end{align}
	
	where 
	$\alpha_I^2= ( \tilde{Q_I}^2  + \tilde{P_I}^2 ), \alpha_S^2 = ( \tilde{Q_S}^2 + \tilde{P_S}^2 ), \tilde{Q_I} = \alpha_I\cos\theta_I, \tilde{P_I} = \alpha_I\sin\theta_I, \tilde{Q_S} = \alpha_S\cos\theta_S, \tilde{P_S} = \alpha_S\sin\theta_S $, where 	$\tilde{Q_I} =\frac{Q_I}{\sqrt{mn-c_1^2}}, \tilde{P_I} =\frac{P_I}{\sqrt{mn-c_2^2}}, \tilde{Q_S} =\frac{Q_S}{\sqrt{mn-c_1^2}}, \tilde{P_S} =\frac{P_S}{\sqrt{mn-c_2^2}}$, and
	$c^2 = c_1^2 +c_2^2, c_1=c \cos\phi, c_2=c \sin\phi $, and therefore, $\alpha_I,\alpha_S,c\geq 0$ and 
	$ 1 \ge (\cos\phi \cos\theta_I \cos\theta_S + \sin\phi \sin\theta_I \sin\theta_S) \ge -1$. 
	As $\phi$ is a fixed angle, for a given correlation matrix, we have the liberty to vary $\theta_I$ and $\theta_S$ to maximize the Bell function given in Eq. \eqref{bell_func_deff}. We therefore find out the intermediatory maximized Bell function as

	\begin{align}\label{bell_func_m}
		B_{m} &= \frac{1}{4 \sqrt{\det(V )} } \big[1 + \exp\left(-\frac{1}{2} m \alpha_I^2 \right) \\
		&+ \exp\left(-\frac{1}{2} n \alpha_S^2 \right) -\exp \left(-\frac{1}{2} \left[ m \alpha_I^2  +n \alpha_S^2 +2 \tilde{c} \alpha_I \alpha_S  \right]\right)  \big],  \nonumber
	\end{align}

	where $\tilde{c} = \max[|c_1|,|c_2|]$.  It can be shown that $B\leq B_m$ straightforwardly at any circumstances. Extreme points of $B_m$ will be determined for $\alpha_S = \sqrt{\frac{m}{n}} \alpha_I$. In fact, $B_m$ shows symmetry $B(\alpha_I,\alpha_S) = B(\sqrt{\frac{n}{m}} \alpha_S,\sqrt{\frac{m}{n}} \alpha_I)$, and such combinations has already been in use before for TMSV states \cite{Banaszek_TMSV_bell, Banaszek_TMSV_bell_PRL}. Expressing everything in terms of $\alpha_I$ as $B_{m} = \frac{M}{4 \sqrt{\det(V)} } $ where $M = \left[1 + 2 A^{-m} -  A^{-(2m  +2 \tilde{c}\sqrt{\frac{m}{n}} ) } \right]$ where $ A = \exp\left(\frac{1}{2} \alpha_I^2 \right)$, therefore $ 0\le\alpha_I\le\infty \to 1 \le A\le \infty$. $\partial_A M = 0$ gives 
	
	\begin{equation}
		A = \left(\frac{m}{m+\tilde{c}\sqrt{m/n}}\right)^{-\frac{1}{m+2\tilde{c}\sqrt{m/n}}}
	\end{equation}
	
	which gives 
	
	\begin{equation}\label{nonlocality_formula}
		B_{max} = \frac{1/4}{ \sqrt{\det(V )} } \left[1 + \left(\frac{ \sqrt{nm}    }{ \sqrt{ nm } +\tilde{c} }\right)^{\frac{    \sqrt{ nm }   }{ \sqrt{ nm } +2\tilde{c} }} \left( \frac{ \sqrt{ n m } +2\tilde{c} }{  \sqrt{ n m } +\tilde{c} }\right)   \right]
	\end{equation}
	
	One can also check $\partial^2_A M < 0$, which ensures maxima. The thermal dissipation of TMSV in the results of \cite{TMSV_nonlocal_Bell_Wigner, mypaper_TMSV_filter} can be recovered from the expression of Eq. \eqref{nonlocality_formula}, with a suitable choice of matrix elements of Eq. \eqref{corr_mat_transformed}. For example, in case of TMSV, by fixing $m=n=\cosh 2r, \tilde{c} = \sinh 2r$, where $r$ is the squeezing factor, one can see that when $r\to \infty, B_{max} \to 1 + \frac{3}{2^{4/3}} = 2.19055$; the state becomes maximally nonlocal \cite{Banaszek_TMSV_bell}. However, in the case of the generalized formulation given in Eq. \eqref{nonlocality_formula}, the non-locality condition:
	$B_{max}\geq 2$ leads to
	
	\begin{align}\label{nonlocality_criteria}
		\frac{1}{16} \left[1 + \left(\frac{ 1    }{ 1 + x }\right)^{\frac{  1   }{ 1 +2 x }} \left( \frac{ 1 +2x }{ 1 + x }\right)   \right]^2 \geq 4  (nm - c_1^2 ) (nm - c_2^2)
	\end{align}
	
	where $x = \tilde{c}/\sqrt{nm} \leq 1$ which can easily be justified from the Schr\"odinger-Robertson uncertainity principle \cite{Robertson_Schrodinger}.

	Note that we accept this form of the violation of the Bell inequality because the state is described by a positive Wigner function \cite{Banaszek_TMSV_bell, Banaszek_TMSV_bell_PRL}. The local operations during LLUBO do not change the maximized Bell function and the violation of local realism, given in Eq. \eqref{nonlocality_criteria}.

	\section{Comparing Entanglement and Nonlocality}
	
	Simon's sufficient criteria for entanglement is  \cite{Simon}
	
	\begin{align}\label{Entanglement_Simon}
		4  ( nm -c_1^2) ( nm -c_2^2) \leq (n^2+m^2) +2|c_1c_2| -1/4 
	\end{align}
	
	The condition furthermore quantifies entanglement in terms of logarithmic negativity. However, we intend to compare here the sufficient conditions of entanglement and nonlocality. 
	The LHS of the Eq. \eqref{nonlocality_criteria} furthermore, can be limited to
	
	\begin{align}
		%	(n^2+m^2) +2|c_1c_2| -1/4 \geq 2nm +2|c_1c_2| -1/4  \geq \frac{4nm}{4}(2 +2|c_1c_2|/nm -1/4nm ) \geq \frac{1}{4}(1 +2|c_1c_2|/nm ) \text{ as } n,m\geq 1/2 \\
		\frac{1}{16} \left[1 + \left(\frac{ 1 }{ 1 + x }\right)^{\frac{ 1 }{ 1 +2 x }} \left( \frac{ 1 +2x }{ 1 + x }\right)   \right]^2 = & \frac{1}{4} (1+x^2) - O[x^3]  \nonumber \\
		\leq \frac{1}{4} (1+x^2) \,\, \text{ in the range } & 0\leq x \leq 1
	\end{align}
	
	%Without loss of generality, we assume $ \tilde{c} = |c_1|$ and $ \tilde{c}' = |c_2|$
	
	As $(n^2+m^2) \geq 2nm$, therefore, if we can prove
	
	\begin{equation} \label{ent_nonlocal_relation}
		nm ( 2 + 2 xx' -\frac{1}{4nm} )  \geq \frac{1}{4} (1+x^2),\, \text{where} \, x' = \frac{\min[|c_1|,|c_2|]}{\sqrt{nm}} , 
	\end{equation}
	
	it will be ensured that non-locality is a sufficient criterion for all Gaussian states to be declared as entangled. However, this does \textit{not} ensure that if the states are entangled, they are non-locally relizable.

	\textbf{Proof:}
	
	From the realization of the mixedness of the state given in Eq. \eqref{mixedness}, we have 
	
	\begin{align}
		& 0\leq x'^2 \leq  1 - \frac{1}{16(nm)^2 (1-x^2)} \nonumber \\	
		& \frac{1}{4(nm) }  \leq \sqrt{(1-x^2)}
	\end{align}
	
	Applying it on LHS-RHS of the Eq. \eqref{ent_nonlocal_relation}, we have
	
	\begin{align}
		&(2 + 2 xx' -\frac{1}{4nm} ) - (\frac{1}{4nm} + \frac{x^2}{4nm} )\geq	2 - \frac{1}{4nm} (2+x^2) \nonumber \\
		&\geq 2 - (2+x^2) \sqrt{(1-x^2)} \geq 0  \,\, \text{in the range} \, 0\leq x \leq 1
	\end{align}
	
	which proves the inequality in Eq. \eqref{ent_nonlocal_relation}. The result justifies that the realization of non-locality is sufficient to witness entanglement in the case of bipartite Gaussian states, but the reverse may not be true. The phenomenon has also been seen in \cite{mypaper_TMSV_filter_STD} before, where the nonlocality remains limited within a small region of entanglement, justifying that entanglement is necessary but not sufficient to realize non-locality.

	\section{Summery} %CONCLUSION
	
	We have formulated maximally optimized Bell's function % as a measure of non-locality
	of bipartite CV Gaussian states in terms of correlation matrix elements, using Wigner phase-space formalism, which eventually offers a boundary condition for local realism. Our analysis also shows that if the bipartite Gaussian state is nonlocally measurable, it is sufficient to prove the state to be entangled. Our analysis, moreover, builds up a connection between entanglement and nonlocality in the case of Gaussian states.

	\begin{acknowledgments}
		The work has partly been supported by the European Commission, MSCA GA no 101065991 (SingletSQL).
		
	\end{acknowledgments}
	
	%\section*{DATA AVAILABILITY}
	%No data has been generated.
	
	%\section*{Conflict of interest}
	%The author has no conflicts to disclose.
	
	%\appendix
	
	%\section{Optimization of Maximally Squeezd Quadrature}\label{optimization_squeezing}

	\nocite{*}

	\bibliography{apssamp}% Produces the bibliography via BibTeX.

\end{document}